\documentclass[super]{rsc-layout}
\usepackage{graphicx,amsmath,amssymb,textcomp}
\graphicspath{{figures/}}

\newcommand\eq[1]{Eq.~\eqref{eq:#1}}
\newcommand\fig[1]{Fig.~\ref{fig:#1}}

\renewcommand\vec[1]{\textrm{\bfseries #1}}
\newcommand\dotprod{\boldsymbol{\cdot}}
\newcommand\diff{\mathrm{d}}

\makeatletter
\newcommand\dash{\penalty\@M-\hskip\z@skip}
\makeatother

\begin{document}

% Running head
% \pagestyle{myheadings}
% \markboth{Horton, H{\"o}f{}ling, R{\"a}dler, and Franosch}{Anomalous diffusion develops among crowding proteins}

% \titlehead{\sffamily\small LMU-ASC 32/09}
%\ohead{LMU-ASC 32/09}
\title{Development of anomalous diffusion among crowding proteins} %bound to a supported lipid bilayer}
\author{%
Margaret R. Horton\footnotemark[1]%
{\renewcommand\thefootnote{\fnsymbol{footnote}}%
\footnotemark[1]\footnotemark[2]\footnotemark[3]}%
,
Felix H{\"o}f{}ling\footnotemark[1]\footnotemark[2]\footnotemark[3]%
{\renewcommand\thefootnote{\fnsymbol{footnote}}%
\footnotemark[3]}%
,
Joachim O. R{\"a}dler\footnotemark[1]%
, and
Thomas Franosch\footnotemark[1]\footnotemark[2]\footnotemark[4]%
}

% \date{}

% abstract: 200 words max
\abstract{%
In cell membranes, proteins and lipids diffuse in a highly crowded and heterogeneous landscape, where
aggregates and dense domains of proteins or lipids obstruct the path of diffusing molecules.
In general, hindered motion gives rise to anomalous transport, though the nature of the onset of this behavior is still under debate and difficult to investigate experimentally.
Here, we present a systematic study where proteins bound to supported lipid membranes diffuse freely in two dimensions, but are increasingly hindered by the presence of other like proteins.
In our model system, the surface coverage of the protein avidin on the lipid bilayer is well controlled by varying the concentration of biotinylated lipid anchors.
Using fluorescence correlation spectroscopy (FCS), we measure the time correlation function over long times and convert it to the mean-square displacement of the diffusing proteins.
Our approach allows for high precision data and a clear distinction between anomalous and normal diffusion.
It enables us to investigate the onset of anomalous diffusion, which takes place when the area coverage of membrane proteins increases beyond approximately 5\%.
This transition region exhibits pronounced spatial heterogeneities.
Increasing the packing fraction further, transport becomes more and more anomalous, manifested in a decrease of the exponent of subdiffusion.}


\maketitle

% footnotes must appear after maketitle
% affiliations first, use letters
\bgroup
\renewcommand\thefootnote{\alph{footnote}}
\footnotetext[1]{%
Center for NanoScience (CeNS), Ludwig-Maximilians-Universit{\"a}t M{\"u}nchen, Geschwister\dash{}Scholl-Platz~1,
80539 Munich, Germany}
\footnotetext[2]{%
Arnold Sommerfeld Center for Theoretical Physics, Fakult{\"a}t f{\"u}r Physik, Ludwig-Maximilians-Universit{\"a}t M{\"u}nchen, Theresienstra{\ss}e~37, 80333 Munich, Germany}
\footnotetext[3]{%
Rudolf Peierls Centre for Theoretical Physics, 1 Keble Road, Oxford OX1 3NP, England, United Kingdom}
\footnotetext[4]{%
Institut f\"ur Theoretische Physik, Universit\"at Erlangen-N\"urnberg, Staudtstra{\ss}e~7, 91058, Erlangen, Germany}
\egroup

% notes follow the affiliations, use symbols
\footnotetext[1]{present address: Unnastr. 48, 20245 Hamburg, Germany}
\footnotetext[2]{Correspondence: mhorton@alum.mit.edu}
\footnotetext[3]{M.\ R.\ Horton and F.\ H\"of{}ling contributed equally to this work.}
\setcounter{footnote}{3}

\section*{Introduction}

Anomalous transport of proteins in a crowded environment is best discussed in real space, in terms of the mean-square displacement (MSD). There, deviations from Fickian diffusion can clearly be inferred in a double-logarithmic representation.
While computer simulations yield the MSD directly, single-particle tracking (SPT)
is essentially the only experimental method that gives access to this quantity~\cite{Greenleaf:2007,Kirstein:2007,Kusumi:2005,Saxton:1997}.
Following the trajectories of individual, fluorescently labeled proteins prompts our spatial imagination and facilitates the discussion of the underlying mechanisms.
Such methods, however, require sophisticated and costly image processing, their temporal resolution is typically limited to fractions of a second, and long-time measurements suffer from bleaching dye due to the intense laser light.

A complementary technique is fluorescence correlation spectroscopy (FCS), which has been established to precisely quantify normal protein diffusion at the micro- and millisecond scales~\cite{Chiantia:2009,Krichevsky:2002,RiglerElson:FCS,Schwille:1999a}.
Compared to SPT, this technique is much simpler and less time consuming in terms of performance and data processing, its temporal resolution is essentially limited by the
responsiveness of the dye, and the laser power is so low that bleaching is greatly reduced.
In particular, FCS offers time windows of several decades and is thus a key technique to study \emph{anomalous} diffusion,
e.g., in the cytoplasm~\cite{Guigas:2007,Wachsmuth:2000}, in cell membranes~\cite{Schwille:1999,Weiss:2003}, and in protein solutions imitating crowding conditions~\cite{Sanabria:2007,Banks:2005,Weiss:2004}.
The central quantity in FCS is the time-correlation function of the fluctuating intensity of the fluorescent light from labeled molecules. A discussion of such a correlation function is more subtle than for plain trajectories, and it requires certain assumptions on the measurement process itself and in many cases even on the nature of the dynamic processes inside the probe.
While there is little ambiguity in the case of dilute systems showing normal diffusion,
the investigation of anomalous diffusion is complicated by such prerequisites.
Here, we convert the FCS autocorrelation function directly to an MSD, from which anomalous diffusion can be easily inferred without resorting to a particular diffusion model \emph{a priori}.
Further, such an analysis could even capture a possible crossover in the MSD from subdiffusion  at short times to normal diffusion at long times.

Membrane proteins are naturally exposed to a highly crowded landscape,
with stark consequences on their transport dynamics.
The cell membrane consists of a heterogeneous lipid bilayer densely packed with integral and peripheral proteins; in addition, some of the lipids are organized into microdomains, and some of the proteins are tethered to the cytoskeleton~\cite{Engelman:2005}.
Measurement of protein transport in cell membranes is thus experimentally challenging.
A pronounced reduction of lateral protein diffusivity has been established by numerous experiments using SPT~\cite{Crane:2008, Kusumi:2005, Vrljic:2002, Tomishige:1998, Feder:1996, Smith:1999, Sheets:1997, Ghosh:1994},
FCS~\cite{Gielen:2005, Lenne:2006, Weiss:2003, Schwille:1999}, or fluorescence recovery after photobleaching~\cite{Kenworthy:2004,Feder:1996}. There is general agreement that this effect is caused by macromolecular crowding~\cite{Zhou:2008, Dix:2008, Kusumi:2005, Ellis:2001a, Saxton:1997}.
Many of the experimental findings have been interpreted in terms of anomalous diffusion~\cite{Kusumi:2005, Smith:1999,  Sheets:1997, Ghosh:1994, Gielen:2005, Weiss:2003, Schwille:1999},
but several experiments also display simple Brownian motion~\cite{Crane:2008, Vrljic:2002, Sheets:1997}.
It was emphasized that anomalous transport should not be considered a universal feature of crowded systems~\cite{Dix:2008},
and it has been noted elsewhere~\cite{Kusumi:2005} that the transport characteristics may depend on the time scale of the observation---which may resolve the controversy.
% Recent experimental evidence suggests that physical crowding and the intrinsic composition of the membrane, rather than its interaction with cytoskeletal components, accounts for the slow diffusion~\cite{Frick:2007}.
Nevertheless,
% measurements of anomalous transport of membrane proteins are experimentally challenging and
the relevant crossover time scales and, more generally, the source of the anomalous behavior have not been identified so far.

To shed light on the origin of anomalous transport in membranes, experimental model membrane systems with an adjustable
degree of crowding are desirable, allowing isolated effects to be studied systematically.
The hope is to observe both the onset of anomalous behavior as well as the crossover to normal diffusion at long times as functions of the system parameters.
Promising candidates for an experimental realization are synthetic lipid bilayers, which offer the advantage that the composition can be controlled precisely.
Such model membranes have provided insight into how diffusion in membranes is affected by immobilized lipids~\cite{Deverall:2005} and lipid phase separation~\cite{Hac:2005}.
Yet, the experimental scenario of physical molecular crowding in membranes remains relatively under-investigated~\cite{Zhou:2008,Melo:2006}.

\begin{figure}
   \begin{center}
      \includegraphics*[width=.8\linewidth]{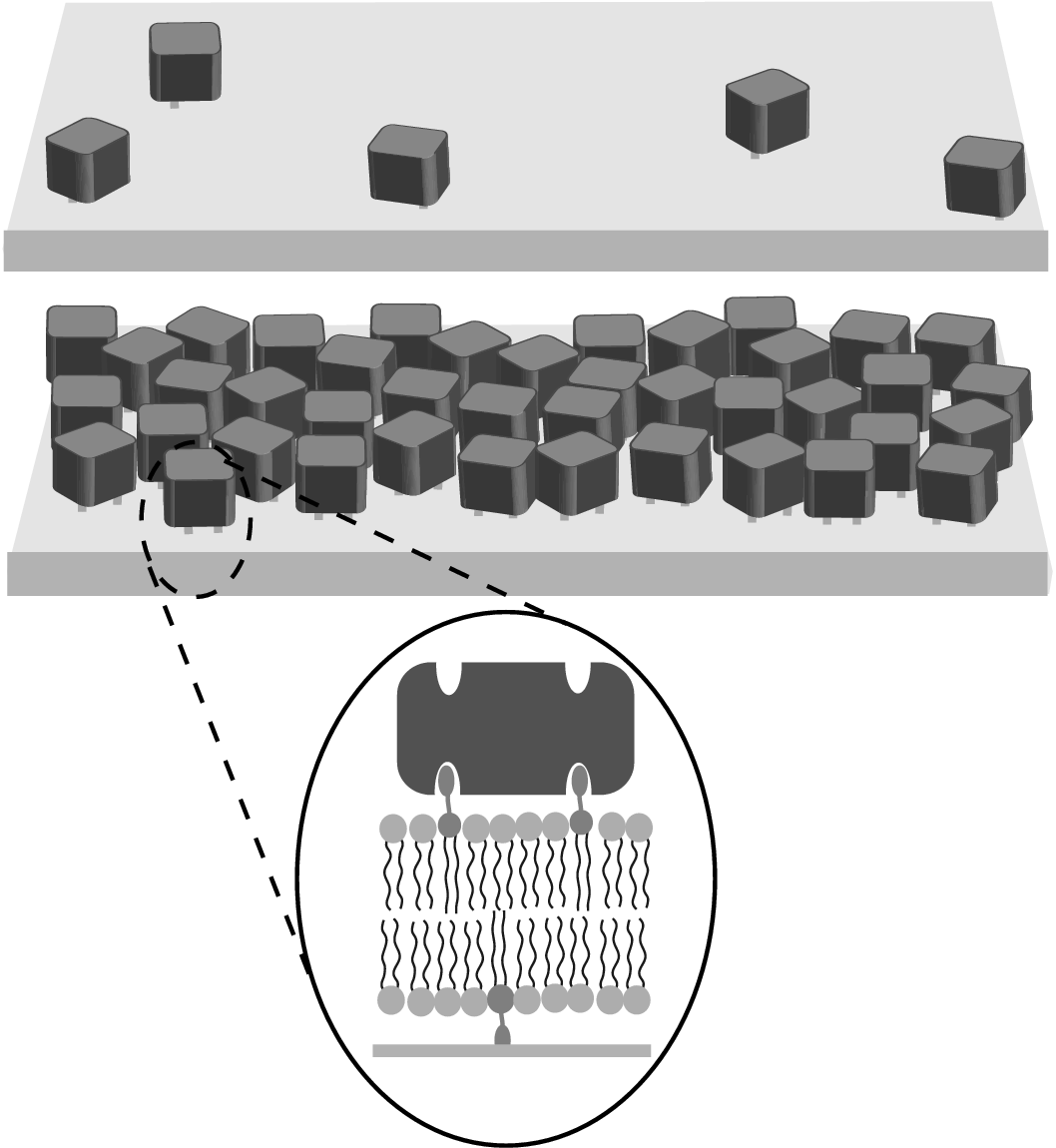}
      \caption{Experimental system for studying protein crowding on a lipid bilayer
membrane. A lipid bilayer of SOPC and biotin-X-DPPE, acting as lipid anchors, is formed on a glass substrate;
avidin binds irreversibly to these anchors. Thus, the surface
coverage of avidin is determined by the concentration of lipid anchors in the
bilayer. A fluorescently-labeled analog of avidin is included in the avidin mixture for
fluorescence correlations spectroscopy (FCS).}
      \label{fig:3dview}
   \end{center}
\end{figure}

In the present work, we use FCS to investigate the systematic development of anomalous transport of proteins bound to a crowded model membrane on a planar substrate.
Specifically, we have studied the motion of the protein avidin in supported lipid bilayers as  function of the protein content (\fig{3dview}).
By exploring a wide range of avidin surface coverages, this system can be fully characterized
in the dilute, intermediate, and crowded regimes,
and drastic changes in the protein diffusivity are observed.
The protein motion is clearly inferred from the obtained MSDs, which permit a clear distinction between normal and anomalous diffusion.

\subsection*{Experimental model system}

Our experimental model system consists of a single lipid bilayer of SOPC containing a fraction of biotin-X-DPPE, which is supported by a transparent glass substrate~\cite{Horton:2007}.
The protein avidin irreversibly binds the
biotinylated lipid anchors and has confined mobility in two dimensions on the surface of
the bilayer. Avidin has two biotin-binding sites that are simultaneously available~\cite{Rosano:1999}. We control the surface coverage of avidin by varying the
concentration of biotinylated lipid anchors in the bilayer: as the concentration of lipid
anchors  is increased, the surface coverage of avidin increases.

We take advantage of the non-crystalline nature of avidin at neutral pH to directly measure how excluded volume affects the protein transport.
Previous characterization of avidin diffusion using fluorescent bleaching techniques showed that at high surface
coverages, avidin does not diffuse on  time and length scales comparable to the lipids~\cite{Horton:2007,Lou:2007}.
In the same lipid bilayer system, the diffusion of a lipid tracer below a layer of avidin protein is normal~\cite{Horton:2007}; however, the diffusion of avidin itself is difficult to quantify with fluorescent bleaching.

% turn off preprint number
\ohead{}

\section*{Theory}

\subsection*{Anomalous transport}

% introduce anomalous transport
The simplest dynamic quantity  characterizing  the motion of molecules is the mean-square displacement (MSD) $\delta r^2(t):=\langle|\vec R(t)-\vec R(0)|^2\rangle$ as a function of the time lag~$t$. Simple Fickian diffusion is quantified by the diffusion constant $D$, and the MSD grows linearly with time, $\delta r^2(t) = 4Dt$ for diffusion in a flat membrane.
Hindered transport is manifested either by a reduction of the diffusion constant or by a modified \emph{shape} of the function $\delta r^2(t)$; the latter case is termed anomalous transport.
One generic scenario is a power-law relation referred to as subdiffusion, $\delta r^2(t) \propto t^\alpha$ and $\alpha<1$, which yields a straight line in a double-logarithmic representation. The characteristic exponent $\alpha$ is then directly read off as the slope of this line.

Power-law behavior in dynamic correlations is mathematically equivalent to the presence of processes with power-law distributed characteristic time scales~\cite{Bouchaud:1990}.
Generically there is always a slowest process, and the subdiffusive behavior is cut off for sufficiently long observation times and crosses over to simple diffusion~\cite{Kammerer:2008,Hoefling:2007,Kusumi:2005}.
In the case of cell membranes, a power-law distribution of characteristic time scales requires a hierarchy of slow processes that hinder the molecular motion. One possible source is trapping by specific binding~\cite{Dix:2008,Saxton:2007}, but the origin of the hierarchical ordering is not obvious here. Another possibility is that time scales are inherited from a spatially heterogeneous structure, which stems either from the presence of very differently sized macromolecular complexes, from self-organized criticality, or which is related to a critical phenomenon---as observed for randomly distributed obstacles~\cite{Saxton:1994,Sung:2006,Hoefling:2006,Hoefling:2008}.

\subsection*{The FCS autocorrelation function}

The FCS method consists of autocorrelating the intensity fluctuations  of fluorescent light from a steadily  illuminated volume. Since the fluorescence is proportional the incident laser intensity $W(\vec{r})$ and the  concentration of labeled molecules $c(\vec{r},t)$ fluctuating in time and space, the collected light in the detector
is determined by
\begin{equation}
I(t) \propto \int\! W(\vec{r})  c(\vec{r},t)\diff^2 r \, ,
\end{equation}
where we consider the molecules confined to a plane. The quantity measured in an FCS experiment is simply the time-autocorrelation function of the fluctuating intensity $\delta I(t) = I(t) - \langle I \rangle$ around the mean intensity,
\begin{equation}
 G(t) = \frac{\langle \delta I(t) \delta I(0) \rangle}{ \langle I \rangle^2} \, ,
\end{equation}
where the normalization is convention. The intensity fluctuations  can be expressed as a weighted average of concentration fluctuations of the labeled molecules.
It is straightforward to adapt the derivation of Ref.~\citenum{BernePecora:DynamicLightScattering}  to two-dimensional systems. Then
the FCS signal
\begin{equation}
 G(t) = \frac{ \int\! \diff^2 r \diff^2 r'\, W(\vec{r}) W(\vec{r}') S(\vec{r}-\vec{r}',t)}{ \langle c \rangle \left[\diff^2 r W(\vec{r}) \right]^2} \,
\end{equation}
is related to the van-Hove correlation function
\begin{equation}
 S(\vec{r}-\vec{r}',t) = \frac{1}{\langle c \rangle} \langle \delta c(\vec{r},t) \delta c(\vec{r}',0) \rangle \, .
\end{equation}
Its spatial Fourier transform is known as the  intermediate scattering function~\cite{Hansen:SimpleLiquids}
\begin{equation}
S(\vec{q},t) = \int\! S(\vec{r}-\vec{r}',t) \text{e}^{\text{i} \vec{q} \dotprod (\vec{r}-\vec{r}')}\diff^2 r\,
\end{equation}
and the FCS autocorrelation can be expressed as
\begin{equation}
 G(t) = \frac{1}{N} \frac{\int\! \diff^2 q\, |W(\vec{q})|^2 S(\vec{q},t)}{\int \diff^2 q |W(\vec{q})|^2} \,
\end{equation}
where $W(\vec{q})$ is the spatial Fourier transform of the intensity profile $W(\vec{r})$ and
\begin{equation}
 N = \langle c \rangle \frac{ (2\pi)^2 |W(\vec{q}=0)|^2 }{\int\! \diff^2 q\, |W(\vec{q})|^2 }
= \langle c \rangle \frac{ \left[ \int\! \diff^2 r\, W(\vec{r}) \right]^2}{ \int\! \diff^2 r\, W(\vec{r})^2 } \, ,
\end{equation}
can be interpreted as the effective number of particles illuminated. For a Gaussian profile of the laser beam
\begin{equation}
 W(\vec{r}) \propto \exp\left(-2 \vec{r}^2/w^2\right) \, ,
\end{equation}
where $w$ denotes the beam waist, the filter function $|W(\vec{q})|^2 \propto \exp(-q^2 w^2/4)$ is again Gaussian, and the number of illuminated particles is found as
$N = \langle c \rangle \pi w^2$.

If only a small fraction of molecules is labeled,
the intermediate scattering function reduces to the incoherent scattering function
\begin{equation}
S(\vec{q},t) \approx F(\vec{q},t) =
\left\langle \text{e}^{ \text{i} \vec{q} \dotprod [\vec{R}(t)-\vec{R}(0)] } \right\rangle \, .
\end{equation}
For normal diffusion $F(\vec{q},t) = \exp(- D q^2 t)$ and again for a Gaussian beam the FCS autocorrelation then attains the simple form
\begin{equation}
G(t) = \frac{1}{N} \frac{1}{1+ t/\tau_D} \, ,
\label{eq:fcs_normal}
\end{equation}
where $\tau_D = w^2/4 D$ denotes the dwell time.
Relaxing the assumption of diffusion, the incoherent scattering function
for two\dash{}dimensional motion is given by
$F(\vec{q},t) = \exp\left[- q^2  \delta r^2(t)/4 \right]$
within a Gaussian approximation.
Hence, the motion is characterized by the mean-square
displacement $\delta r^2(t) :=\langle \Delta \vec{R}(t)^2 \rangle$, and the FCS autocorrelation is then given by
\begin{equation}
 G(t) = \frac{1}{N} \frac{1}{1+ \delta r^2(t)/w^2 } \, .
 \label{eq:fcs_msd}
\end{equation}
A variant of this formula for the three-dimensional situation was given during this work by {Shusterman \emph{et al.}~\cite{Shusterman:2008}}; a generalization to polymers is found in Ref.~\citenum{Winkler:2006}.

The fluorescence conversion in general includes  intermediate dark triplet states, which modifies the FCS autocorrelation at short times.  To account for the photo physics, a
time-dependent  factor has been introduced~\cite{Widengren:1995}, yet if the triplet lifetime is much shorter than the time scales of interest, this factor may be safely ignored.
In addition, we assume that the fraction of fluorophores in the triplet state is sufficiently
small and thus negligible for the normalization of $G(t)$.

\section*{Experimental}

\subsection*{Materials}

The lipids 1-stearoyl-2-oleoyl-\textit{sn}-glycero-3-phosphocholine (SOPC, Avanti Polar Lipids, Alabaster, Alabama, USA), N-((6-(biotinoyl) amino)hexanoyl)-1,2-dihexadecanoyl-\textit{sn}-glycero-3\dash{}phospho\-eth\-a\-nol\-amine triethylammonium salt (biotin-X-DPPE, Invitrogen, Karls\-ruhe, Germany) and 2\dash{}(4,4\dash{}difluoro-5-octyl-
4-bora-3a,4a-diaza-\textit{s}-indacene-3-pentanoyl)-1-hexadecanoyl-\textit{sn}-glycero-3-phospho\-cho\-line\linebreak ($\beta$-BODIPY 500/510C\textsubscript{5}-HPC, Invitrogen) were prepared in chloroform. The proteins egg white avidin and fluorescently labeled Alexa Fluor 488-avidin (Alexa488-avidin) and the fluorophore rhodamine 6G were also from Invitrogen. HPLC-grade chloroform, acetone, isopropanol, and ethanol were from Carl Roth (Karlsruhe, Germany). Reagent-grade NH\textsubscript{4}OH, 37\% HCl, and H\textsubscript{2}O\textsubscript{2} were purchased from Sigma (St. Louis, Missouri, USA).

\subsection*{Preparation of the bilayer with bound avidin}
Cover glass slides (24\,mm$\times$24\,mm) were cleaned in a 3-stage process and rinsed with
deionized (DI) water (MilliQ Corp) between each step. First, samples were boiled in a
solution of 5:1 volume ratio of acetone/DI water for 10\,min. Second, samples were boiled
in 1:1:5 DI water/H\textsubscript{2}O\textsubscript{2}/HCl for 15$-$20\,min, rinsed, then boiled in 1:1:5 DI
water/H\textsubscript{2}O\textsubscript{2}/NH\textsubscript{3}OH.
Lipid solutions containing 3\,mg total lipids were mixed in HPLC-grade
chloroform (Roth) in clean glass vials. Then, the chloroform was evaporated from the
vials overnight and then redissolved in HPLC-grade isopropanol to a final concentration
of 1.5\,mg/mL.  This solution was used for spin-coating by dropping 200\,\textmu{}L of solution
onto the clean glass substrates followed by immediate acceleration of the glass to 3000
RPM applied for 60\,s in a Delta~10 spin-coater (BLE Lab Equipment, Radolfzell,
Germany). The lipid-coated substrates were dried for 1\,hr in a vacuum chamber then placed
into Teflon-metal chambers that screw together with an O-ring to create a water-tight seal
to the glass. DI water was added to the chambers to hydrate the lipid film. After overnight
incubation, the substrates were rinsed with DI water and stored at 35$-$40\,\textdegree{}C for 8\,hrs,
allowing the bilayer to anneal and to remove any multi-bilayers on the substrate~\cite{Jensen:2007,Mennicke:2002}.
% hydration of spin-coated lipids at high temperature ensures the removal of multilayered structures.
The complete removal of possible multilayered structures was further verified by
our previous work using x-ray reflectivity and lipid microscopy~\cite{Horton:2007}.
The samples were again rinsed with DI water, then PBS buffer. Lipid
bilayers labeled with 0.005mol\% $\beta$-BODIPY 500/510C\textsubscript{5}-HPC show normal diffusion, even under a fully
crowded layer of avidin. Avidin protein solutions containing 50$-$100\,\textmu{}g total protein were
added to the substrates and stored overnight at 35\,\textdegree{}C. Before measurement with FCS at
room temperature, the substrates were rinsed rigorously with PBS buffer to remove any
non-bound protein. The avidin protein solution contained the fluorescent analog of
Alexa488-avidin with 1:20$-$1:1000 Alexa488/avidin dye ratio. The dye ratio is
optimized at each lipid anchor concentration to maximize the signal-to-noise ratio and
obtain $N\approx 1$.

\subsection*{Fluorescence correlation spectroscopy}
For the FCS measurements, the avidin mixture bound to the membrane contained a small fraction (0.05$-$0.001) of fluorescently-labeled Alexa488-avidin.
ConfoCor2 (Zeiss) with a confocal LSM 510 microscope was used with a 488-nm Ar laser with power
\linebreak
165\,\textmu{}W, 6.1\,mA. Calibration was done with with Rhodamine~6G dissolved in PBS at
concentration of 100nM to determine the dimensions of the focal volume and illuminated
membrane area. The beam waists obtained varied between $w=0.16$\,\textmu{}m and 0.18\,\textmu{}m.
All measurements were performed at 21$-$22\,\textdegree{}C.

To ensure the $z$-stability of the sample for long measurement times, the drift has to be minimized and controlled.
A shift of  0.4\,\textmu{}m  in $z$-direction results in a 50-fold loss in intensity, and such samples
cannot be analyzed. During our measurements, the sample did not drift more than 0.3\,\textmu{}m,
and artifacts in the data due to drift are not expected~\cite{Benda:2003}. We have also measured the signal
from an empty chamber with a bilayer but no protein to determine the background noise
level. The measurements were completed with counts per molecule of 1.1$-$1.8, and a noise
level $<10$\,kHz. Measurements were repeated on at least 2~different samples in at least 4~different spots.

\subsection*{Laser scanning microscopy (LSM)}
For fluorescence imaging of the samples, a large pinhole ($>600$\,\textmu{}m) was selected for planar
imaging of the  bilayer. The laser power for imaging was 6\,mW with pixel time $<2$~\textmu{}s.
The LSM measurements allowed verification that the membranes were intact and free of defects.
The micrographs in \fig{micrographs} show averages over 8 images each.

\subsection*{Statistical analysis}

To determine whether diffusion is normal or anomalous, a least-squares fitting routine
using the experimental standard deviation~\cite{Wohland:2001} was applied. The parameters $N$, $\tau_D$, and, in the case of anomalous diffusion, $\alpha$, were varied, and the autocorrelated data $G_\text{exp}(t)$ were
compared to the diffusion model, $G_\text{model}(t)$, using the Levenberg-Marquardt algorithm in
\textsc{matlab}. The model, normal or anomalous, and the corresponding parameters with the
smallest $\chi^2$ were chosen.

\section*{Results and Discussion}

\begin{figure}
   \begin{center}
      \includegraphics*[width=.9\linewidth]{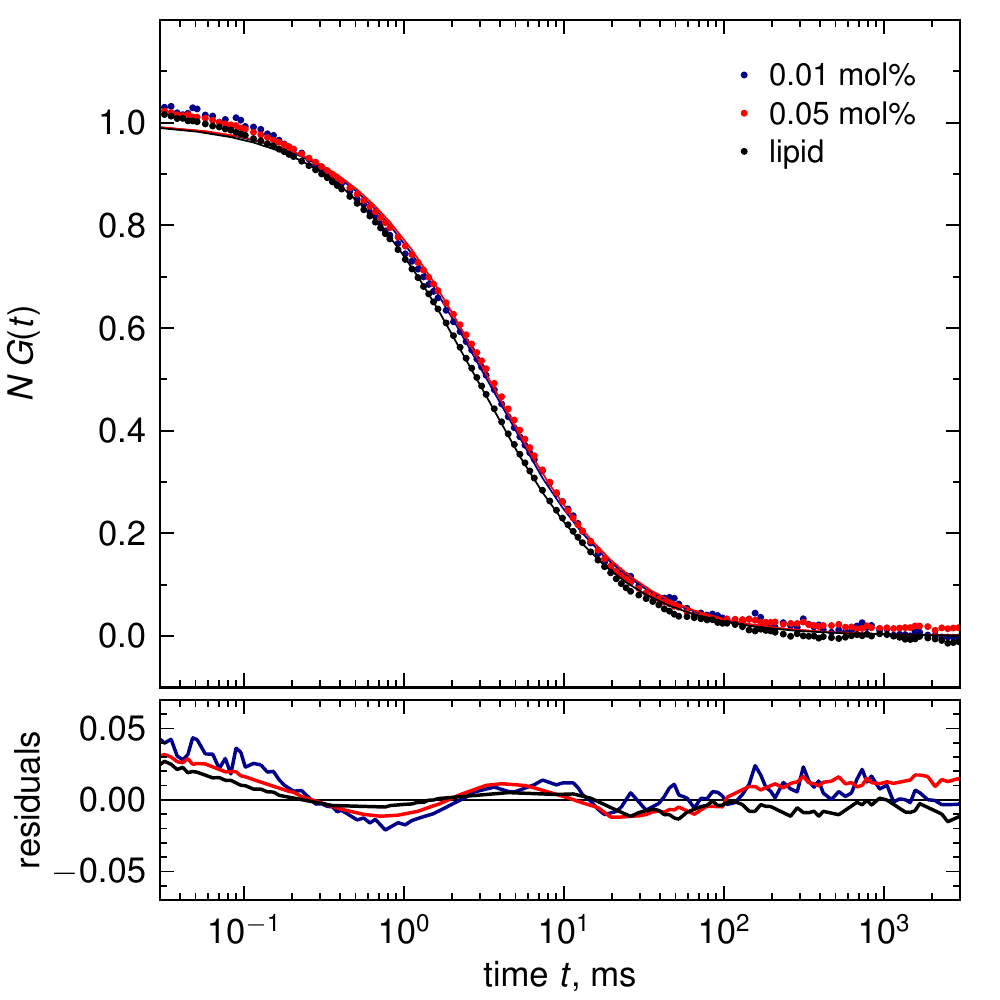}
      \caption{Autocorrelated FCS data of the labeled avidin bound to lipid bilayers containing 0.01 (red) and 0.05 (blue)
mol\% lipid anchors. At such low surface coverages, the data are fitted with the normal diffusion model.
In a separate FCS measurement of labeled lipids (black), normal diffusion comparable to the one of avidin is found.
The inset magnifies the region around the diffusion time.}
      \label{fig:normal_diffusion}
   \end{center}
\end{figure}

\subsection*{Dilute case}
With the goal of studying the dynamics of avidin on the surface of the bilayer over a full
range of surface concentrations, we first examined the dilute case, or the absence of crowding.
A very low surface coverage was obtained by preparing a bilayer containing 0.01$-$0.05\,mol\% biotinylated lipid anchors.
For such low avidin coverage, the measured FCS autocorrelation function $G(t)$ is fit well
by the normal diffusion model for two dimensions, \eq{fcs_normal}, see \fig{normal_diffusion}.
Using $D= w^2/4\tau_D$, the diffusion constant of avidin is
$D= 2.2 \pm 0.1$\,\textmu{}m\textsuperscript{2}/s and
$2.0 \pm 0.3$\,\textmu{}m\textsuperscript{2}/s at 0.01 and 0.05\,mol\% lipid anchors, respectively.
Previous FRAP measurements of the homologous protein streptavidin bound to a similar bilayer
system resulted in comparable values for $D$~\cite{Lou:2007}.
Note that a size-dependence of the diffusion constant, such as that described by the
Saffman-Delbr\"{u}ck diffusion model~\cite{Saffman:1975} and adaptions thereof~\cite{Gambin:2006,Guigas:2006}, is not directly relevant here despite the large size difference between lipid and protein molecules; avidin is not a membrane inclusion, but it is rather bound to the lipid anchors in the bilayer.
Using x-ray reflectivity, we have demonstrated in a previous study~\cite{Horton:2007} that a distinct water layer separates the avidin protein layer from the lipid bilayer.

In a separate control experiment, we have measured the motion of the lipids within the bilayer.
For this FCS measurement, 0.005\,mol\% of lipids were replaced by the
fluorophore $\beta$-BODIPY 500/510C\textsubscript{5}-HPC
as suggested in Ref.~\citenum{Benda:2003}.
We found that the lipids in the bilayer with and without bound avidin diffuse
normally with $D_\text{lipid} = 2.7 \pm 0.4$\,\textmu{}m\textsuperscript{2}/s. This value is similar to previous measurements of the same system using the continuous bleaching method~\cite{Horton:2007}.
Thus, the bound protein has a diffusion constant comparable to the lipids in the underlying bilayer, and we conclude that at low avidin coverage, avidin transport is essentially limited by the diffusion of the lipid anchors. In particular, no signs of anomalous transport were found.
A similar result was obtained in studies of diffusion in surfactant bilayers~\cite{Gambin:2006}.
Using a transbilayer peptide with streptavidin grafted onto one end, diffusion of the
peptide with streptavidin is negligibly slowed compared to the bare peptide lacking
streptavidin.

\subsection*{Anomalous diffusion}

\begin{figure}
   \begin{center}
      \includegraphics*[width=\linewidth]{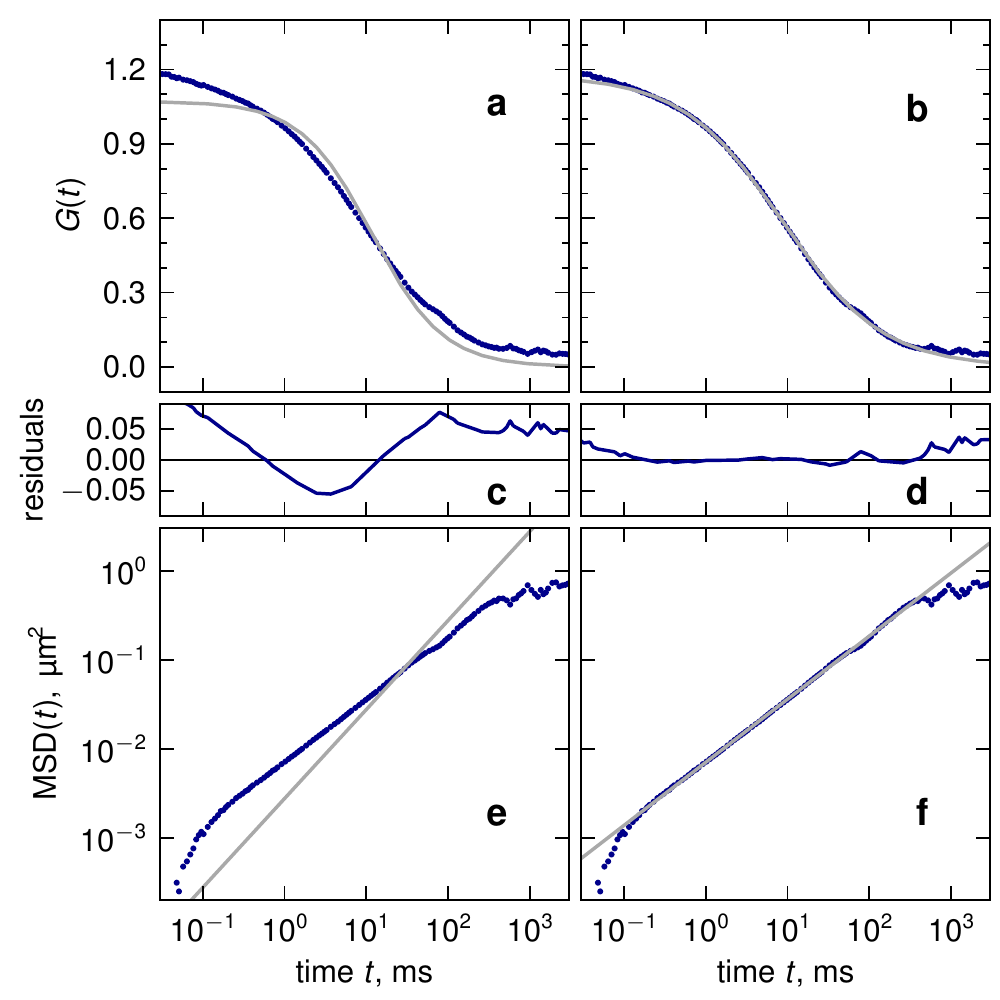}
      \caption{Analysis of the FCS autocorrelation function $G(t)$ at a concentration of 1\,mol\% biotinylated lipid anchors. The data are fitted
much better  with an anomalous diffusion model where $\alpha=0.71$ (panels b,d) rather than with the normal diffusion model (panels a,c).
A double-logarithmic representation of the MSD, obtained via \eq{msd_fcs}, yields a straight line with slope $\alpha$ (panel~f), and is clearly not compatible with normal diffusion, i.e., slope~1 (panel~e).}
      \label{fig:1mol_combo}
   \end{center}
\end{figure}

At sufficiently high concentrations of avidin bound to the lipid bilayer, we expect deviations from normal diffusion.
Increasing the concentration of lipid anchors to 1\,mol\%, we found that the normal diffusion model, \eq{fcs_normal}, produces a poor fit as demonstrated in \fig{1mol_combo}a.
Instead, the data are fitted much better using a model function for anomalous diffusion,
\begin{equation}
 G_{\text{anom}}(t) = \frac{1}{N} \frac{1}{1+(t/\tau_d)^\alpha},
 \label{eq:fcs_anomalous}
\end{equation}
which introduces the exponent $\alpha$;
here, the fit yields $\alpha = 0.71$; see \fig{1mol_combo}b.
Such a fit implies subdiffusive motion, i.e., a MSD growing with a non-integer power in time, $\delta r^2(t)\propto t^\alpha$, which follows by comparison to \eq{fcs_msd} derived
%in the Theory section
within a Gaussian approximation.
An important indicator of the fit quality is to examine the residuals $G(t)-G_\text{model}(t)$,
see \fig{1mol_combo}c,d.
The  residuals for the normal diffusion model %weighted with the experimental error
show a pronounced minimum, which indicates a poor fit~\cite{Schwille:1999}.
Examination of the residuals also suggests that the data are best fit in the time window
0.1\,ms $<t<300$\,ms, and that this window is limited by statistical noise at long times.

Let us comment that a superposition of two normally diffusing components~\cite{Petrov:2008},
\begin{equation}
G_\text{two-comp}(t)=\frac{1}{N}
\left(\frac{N_1/N}{1+t/\tau_{d1}} + \frac{N_2/N}{1+t/\tau_{d2}}\right),
\end{equation}
where $N=N_1+N_2$, also allows for a description of the data with similarly small residuals.
Such a fit, however, is rather susceptible to the precise fit range; changing the upper end of the fit window
from 200\,ms to 3\,s, the ratio $\tau_{d2}/\tau_{d1}$ varies between 14 and 25, while marginally affecting the parameters of the anomalous model, \eq{fcs_anomalous}.
For such large ratios of $\tau_{d1}/\tau_{d2}$, the FCS correlation functions resulting from both models
can hardly be distinguished in the conventional semi-logarithmic representation.
Both models, however, differ significantly in their long-time decay visible on double logarithmic scales.
An unambiguous discrimination in the present case would require a lower noise level in $G(t)$ of 1\%.
Nevertheless we shall not pursue the two-component fit in the following, since the parameters extracted are arbitrary to a large extend
and
 it remains unclear what the nature of the two independently diffusing components
should correspond to.

The usual semi-logarithmic plot of $G(t)$ mainly exhibits the dynamics at the time scale $\tau_d$,
where $G(t)\approx 1/2N$. A characterization of anomalous diffusion, however, requires sensitivity to a broad
range of time scales, ideally spanning several decades. In particular, the behavior at long times is important, although difficult to
access due to the relatively small signal of the autocorrelation function.
Similarly, a least-squares fit is optimized in the vicinity of $\tau_d$ where the statistical noise in $G(t)$ is small~\cite{Wohland:2001}.
There have been important efforts to improve the quantitative evaluation and analysis of
autocorrelated FCS data. The maximum entropy method, for example, is an adaptation
of the $\chi^2$ fitting procedure that considers a distribution of characteristic diffusion times~\cite{Sengupta:2003,Banks:2005}. But as noted
already, purely subdiffusive transport is equivalent to a continuous power-law distribution of characteristic time scales.

\subsection*{Conversion to the mean-square displacement}

The conventional analysis of an FCS measurement is based on model functions that depend on a few parameters such as the diffusion time $\tau_D$ and the number of fluorophores $N$ in the observation volume.
Then the diffusion coefficient can be directly extracted by means of a fitting algorithm.
Such a procedure works as long as the model function $G_\text{model}(t)$ reflects the relevant physics of the sample under investigation.
The selection of an appropriate model function is essential for data fitting and the subsequent interpretation.
If the dynamics of the system under investigation is unknown, the proper choice is sometimes unclear and a subjective issue.
Usually one has to decide \emph{a priori} whether to use a model for normal diffusion, a superposition of diffusing species, or a model for subdiffusion.
For the latter, the situation is even more complicated, since subdiffusion generically crosses over to normal diffusion for sufficiently long times~\cite{benAvraham:DiffusionInFractals, Kammerer:2008, Hoefling:2007}.

Rather than extending existing models by such crossover phenomena, here we suggest converting the autocorrelated FCS data, $G(t)$, directly to the MSD,
\begin{equation}
\delta r^2(t)=w^2\left(\frac{1}{N\, G(t)}-1\right),
\label{eq:msd_fcs}
\end{equation}
simply by inversion of \eq{fcs_msd}.
This eliminates the need to select a model function, permitting an unbiased interpretation of the experiment.
Then, a double-logarithmic representation of the MSD reveals the dynamic features at all scales, which can immediately be identified with the naked eye.
The beam waist $w$ is accessible from independent experiments, and the number~$N$ of labeled molecules can be inferred from the mean intensity directly, or alternatively, from $G(t)$ at short (but not too short) times.
Note that the precise value of $N$ hardly affects the shape of the resulting MSD at intermediate and long time scales, and the value of $w$ merely provides the conversion to absolute length.

The inversion of the FCS autocorrelation resolves the tail at long measurement times, at the price of magnifying the noise for these data. Thus, the experimental challenge is to keep the noise level in $G(t)$ low.
Statistical noise is minimized by collecting more data: longer measurement times give better $G(t)$ statistics.
In the present work, we take advantage of the stability of the system and have measured for relatively long times of approximately 200\,s, significantly improving the statistics at the time scale of 1\,s.
An additional consideration is the fluorescent signal within the focus area. In FCS, the best signal-to-noise ratio for $G(t)$ is obtained by optimizing the concentration of fluorescent molecules such that the number of fluorophores diffusing within the measurement volume is as close to~1 as possible. For the systematic examination of different surface coverages, the fraction of dyed protein thus had to be optimized for each protein concentration.

Applying the data inversion method to the example in \fig{1mol_combo} confirms our earlier conclusion that the
avidin diffusion is anomalous (\fig{1mol_combo}e--f). The quality of the fit for $t>100$\,ms can be more
clearly assessed from visual analysis of the MSD plots compared to studying the $G(t)$
plots of the same data (\fig{1mol_combo}a--b).

% For the graphical FCS data presented in this paper, we have employed the following the data
% analysis protocol: first, a least-squares fit of $G(t)$ in terms of $\chi^2$ criteria using Eq. 4 was
% performed. Then, the estimated parameters $N, \tau_D$, and $\alpha$ were applied in Eq. () to plot the
% MSD. We then examined the MSD plots and readjust $N$ and $\alpha$ manually by less
% than 10\% to give the best overall fit. The plots presented here represent the final fits
% based on examining the MSD.

\begin{figure}
   \begin{center}
      \includegraphics*[width=\linewidth]{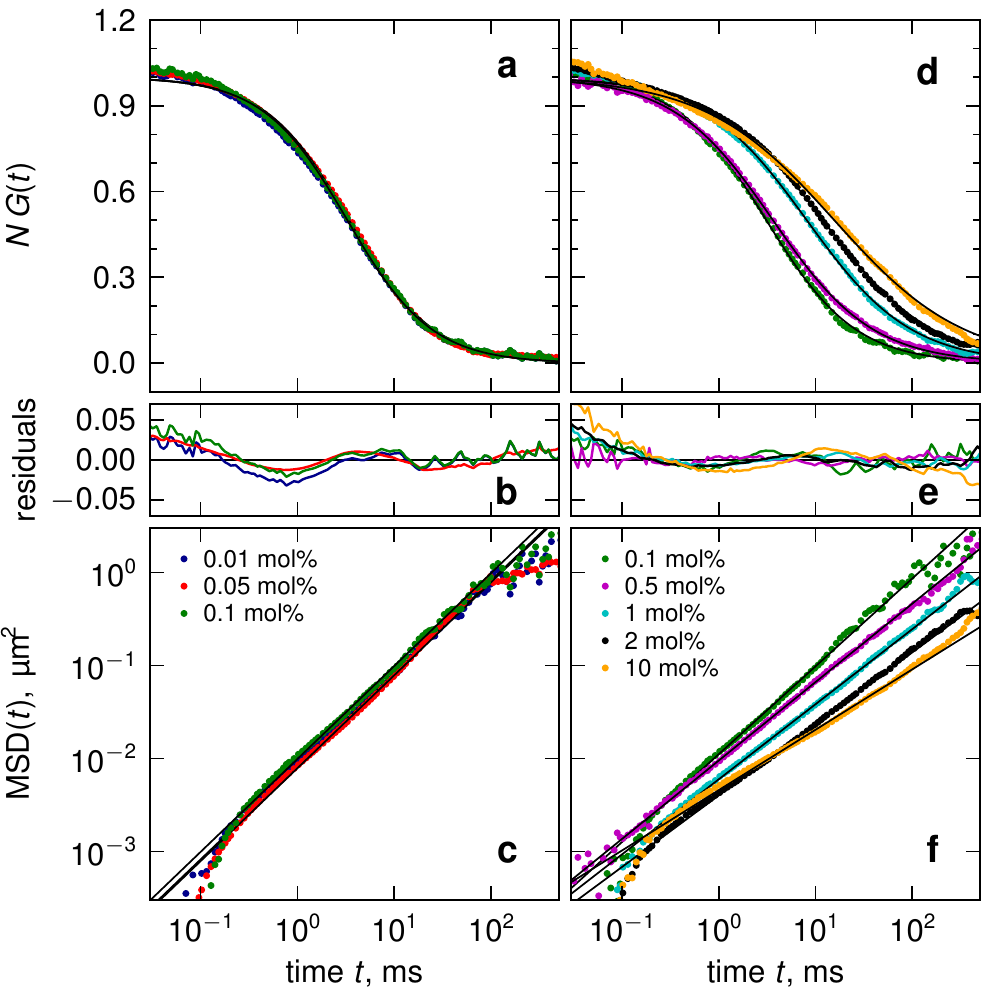}
      \caption{FCS autocorrelations and mean-square displacements obtained via
\eq{msd_fcs} for dilute (panels~a--c) and crowded (panels~d--f) avidin surface coverages.
As the concentration of crowding protein is increased, the slopes of the MSDs decrease in the double-logarithmic representation (panel~f). Thin black lines indicate fits to the FCS data using model functions for normal diffusion (a--c) and subdiffusion (d--f), respectively.}
      \label{fig:msds}
   \end{center}
\end{figure}

\subsection*{Development of the exponent of subdiffusion}

To explore the consequences of crowding on the avidin motion, the concentration of lipid anchors in the bilayer was systematically increased. Several independently
prepared samples were measured in at least four different illumination spots, enabling an estimate of the experimental error for the exponent of subdiffusion.
In the dilute regime with biotinylated lipid anchor concentrations of $0.01-0.05$\,mol\%, the avidin proteins undergo normal diffusion as discussed already, and which is further demonstrated by inspection of the MSD in \fig{msds}c.

\begin{figure}
   \begin{center}
      \includegraphics*[width=.8\linewidth]{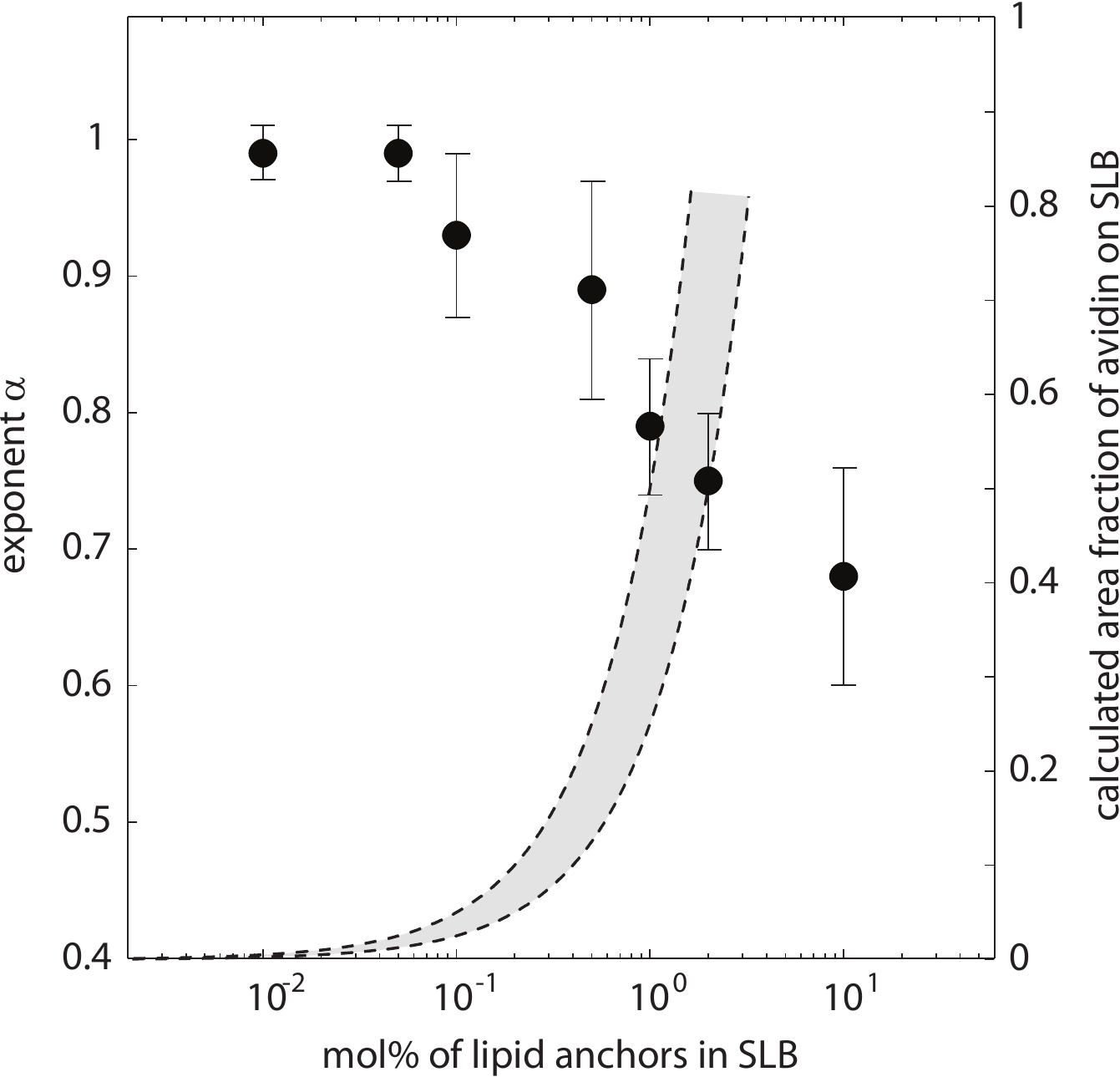}
      \caption{The exponent of subdiffusion~$\alpha$ (symbols, left axis) decreases as a function of the concentration of lipid anchors in the bilayer. The onset of anomalous diffusion occurs at a lipid anchor concentration of 0.1\,mol\%, corresponding to an approximately 5\% surface coverage of protein on the bilayer.
Dashed lines, right axis: Upper and lower estimates for the area fraction of protein covering the bilayer.}
      \label{fig:exponents}
   \end{center}
\end{figure}

At a biotinylated lipid concentration of 0.1\,mol\%, the FCS autocorrelation function starts to deviate from normal diffusion; a satisfactory fit is obtained using the model for anomalous diffusion, \eq{fcs_anomalous}, in \fig{msds}d,e.
As the biotinylated lipid anchor concentration is increased further, avidin transport on the membrane becomes significantly slower, as is apparent from the decreasing slope of the MSDs in a double-logarithmic plot in \fig{msds}f.
Thus, the exponent of subdiffusion $\alpha$ decreases monotonically as more avidin crowds on the surface of the lipid bilayer (\fig{exponents}), and we can quantify the development of anomalous diffusion.
In the most crowded regime, where there is an excess (10\,mol\%) of lipid anchors, we find $\alpha = 0.68 \pm 0.08$.
Interestingly, the obtained MSDs exhibit single power-law behavior in the accessible time window, and the expected crossover to normal diffusion has not yet set in.
In this case,  the conventional method of fitting \eq{fcs_anomalous} to the FCS function and the conversion to the MSD yield equivalent results for the exponent~$\alpha$.

Thus far, we have described the surface coverage of avidin in terms of the
concentration of lipid anchors that we used in preparing the lipid bilayer.
To better depict crowding, the approximate area fraction of the membrane covered with
avidin may be estimated based on the geometry of the proteins and lipids.
We assume that each lipid in the bilayer occupies an area of 50\,\AA\textsuperscript{2} and each avidin molecule  occupies a square of dimension 5\,nm$\times$5\,nm.
Depending on whether the avidin molecules are bound by either one or two biotinylated lipid anchors,
upper and lower estimates of the area fraction covered with avidin follow as indicated in \fig{exponents}.
Note that the other biotin-binding protein streptavidin was shown to bind to a membrane with either one or two biotin anchors~\cite{Lou:2007}.
The onset of anomalous diffusion at 0.1\,mol\% corresponds to a protein area coverage of approximately 3$-$5\%, which is a surprisingly low value.

\begin{figure}
   \begin{center}
      \includegraphics*[width=.9\linewidth]{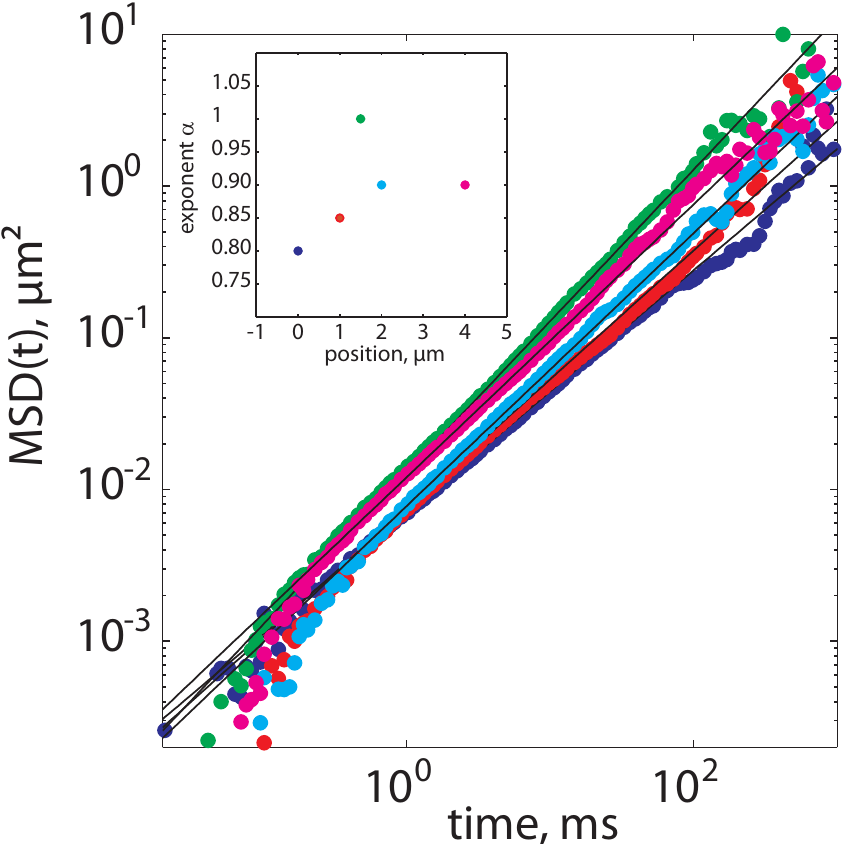}
      \caption{Lateral scanning of a sample with 0.5\,mol\% lipid anchors yields
MSDs exhibiting subdiffusion with position-dependent exponents.
Inset: the exponents as a function of the measurement position.}
      \label{fig:lateral_scanning}
   \end{center}
\end{figure}

\subsection*{The transition regime}
\begin{figure}[t]
   \begin{center}
      \includegraphics*[width=\linewidth]{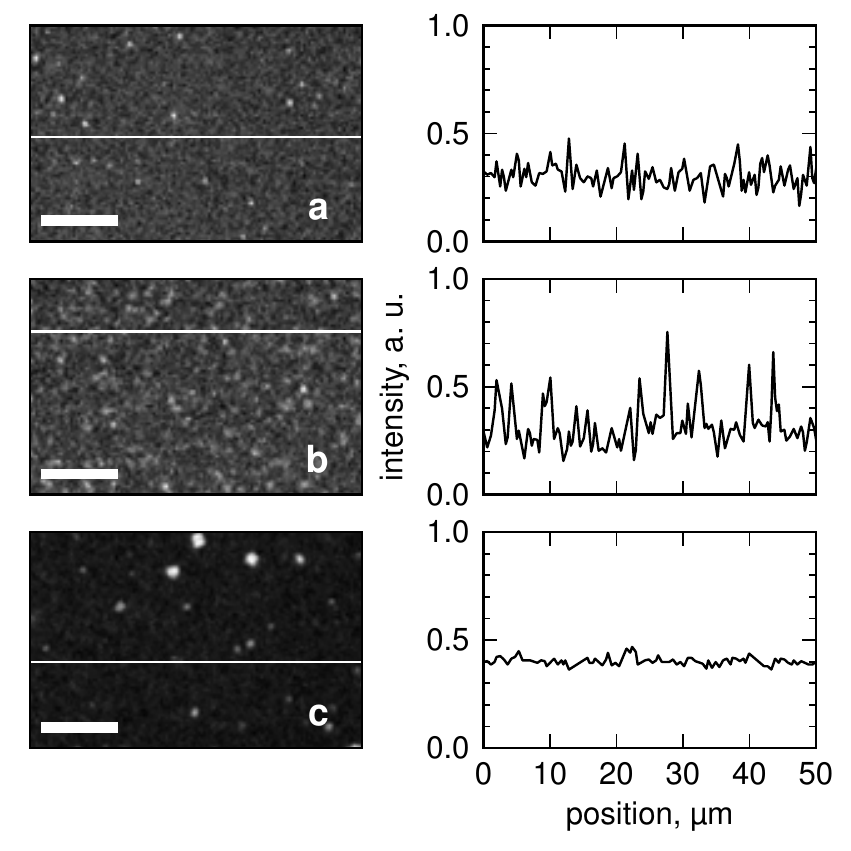}
      \caption{Left: fluorescence micrographs at lipid anchor concentrations of (a) 0.05\,mol\%, (b) 0.5\,mol\%, and (c) 10\,mol\% reveal spatial heterogeneities in the transition regime. Scale bars are 10\,\textmu{}m.
Right: intensity profiles of the corresponding micrographs, curves are offset for clarity.}
      \label{fig:micrographs}
   \end{center}
\end{figure}

By systematically varying the concentration of avidin, we have determined the onset of anomalous diffusion due to molecular crowding. A thorough investigation
of these intermediate concentrations may elucidate  the microscopic origin of the subdiffusive behavior. First, we observe that
 the relative statistical error
in the anomalous diffusion exponent is significantly higher at intermediate lipid anchor concentrations of 0.1 and 0.5\,mol\% (\fig{exponents}). Second,
evidence for heterogeneous structures is found by confocal fluorescence
micrographs in \fig{micrographs}, displaying the distribution of labeled avidin bound to the membrane. The lipid bilayer containing 0.5\,mol\% of biotinylated lipid has a more irregular intensity profile in comparison to the dilute and crowded micrographs with 0.05 mol\% and 10 mol\%, respectively.
The fluorescence micrographs reveal
$1-5$\,\textmu{}m-sized features in the transition regime.

\begin{figure}
   \begin{center}
      \includegraphics*[width=.9\linewidth]{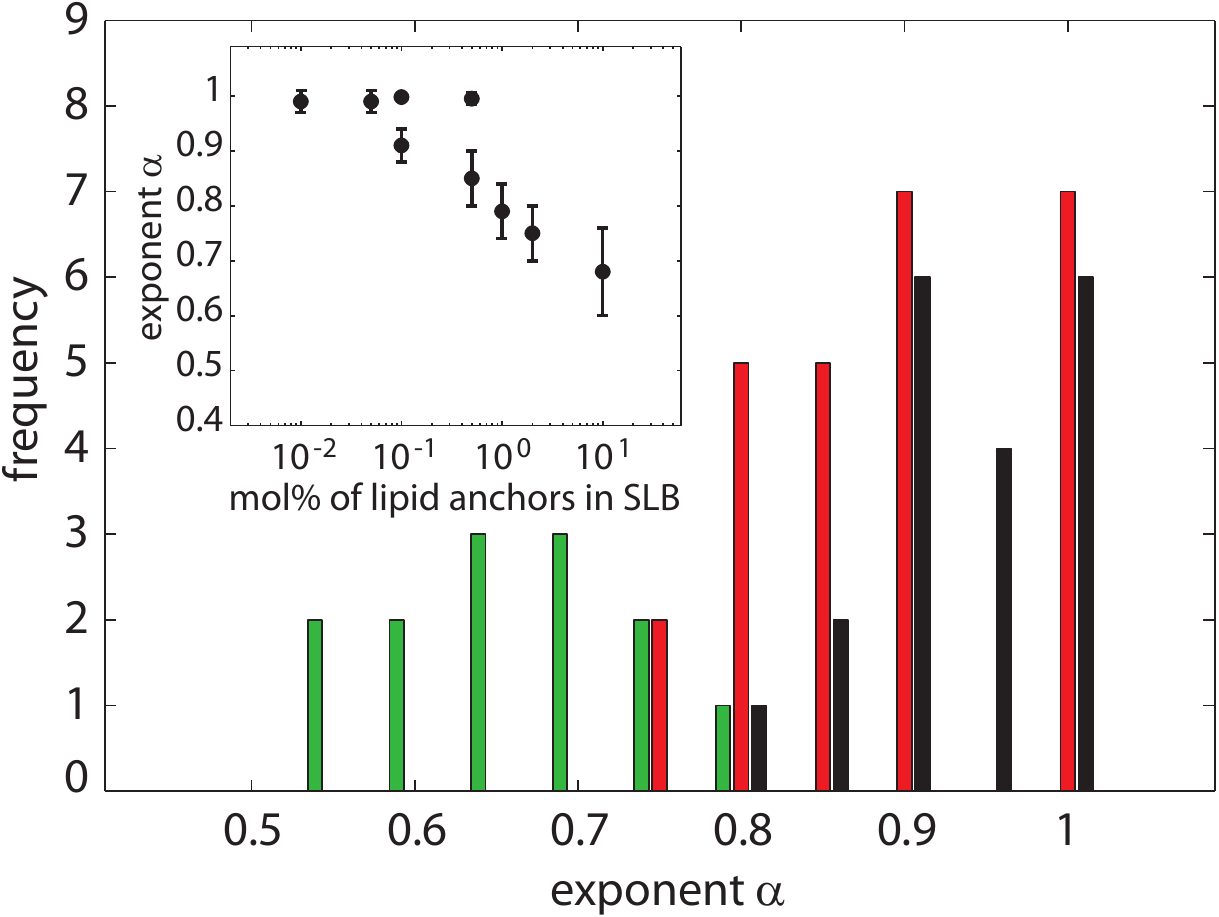}
      \caption{Histograms of the exponent of subdiffusion $\alpha$ from measurements at different positions for lipid anchor
concentrations of 0.1\,mol\% (black), 0.5\,mol\% (red), and 10\,mol\% (green).
Inset: mean exponents as a function of lipid anchor concentration with error bars indicating the variance of the histograms.
In the transition regime, diffusive and subdiffusive spots were treated separately.}
      \label{fig:histograms}
   \end{center}
\end{figure}

These spatial heterogeneities prompted
further spatial FCS characterization of the transition regime.
Probing the dynamics  on various well separated spots
allows autocorrelated FCS data locally to be collected and local mean-square displacements to be extracted, as shown in  \fig{lateral_scanning}. All curves are well represented by
the subdiffusive law, \eq{fcs_anomalous}, where  the exponent interestingly depends on the position. Thus our experiments
reveal large variations in $\alpha$ on a 1\,micron
scale.  Repeated measurement of the same spot after 1\,h showed no changes in  $\alpha$,
corroborating the notion of long-living, spatial heterogeneities.

From the FCS measurements we have collected  histograms of the local exponents $\alpha$ in \fig{histograms}.
The distributions in the values of $\alpha$ suggest qualitatively that there are two classes of transport within the transition regime: there are molecules that appear to
diffuse normally and molecules that show subdiffusive behavior.
The histogram for the 10\,mol\% lipid anchor measurements, however,  appears as a single and very broad distribution.
Consequently the spatial heterogeneities are also present  in the crowded
regime, where at each spot the dynamics is anomalous.
Mean and variance of the histograms are displayed in the inset of \fig{histograms}. Treating the two classes in the transition regime separately
results in significantly smaller error bars compared to \fig{exponents}.

\section*{Conclusions}

% The obtained MSDs allow for a clear distinction between normal and anomalous diffusion over ... from which subdiffusion is inferred over more than two decades in time.

% Understanding the complex membrane transport processes responsible for different cellular membrane processes, such as the assembly of molecular domains or sorting of lipids and proteins, requires diffusion models that account for heterogeneity and crowding. FCS offers the advantage of studying diverse time scales relevant for cell membrane transport processes. Using a stable and well-defined system of membrane proteins crowding on the surface of a lipid bilayer membrane, we experimentally demonstrate a transition from normal to anomalous diffusion. Examining the MSD allows data analysis free of the constraints imposed by fitting the autocorrelated data to a pre-determined model.

% FCS offers the advantage of studying time scales relevant for cell membrane transport processes.
% Using a stable and well-defined system of membrane proteins, crowding on the surface of a lipid bilayer membrane, we experimentally demonstrate a transition from normal to anomalous diffusion.
% Understanding the complex transport dynamics in cell membranes responsible for various cellular membrane processes, such as the assembly of molecular domains or sorting of lipids and proteins, requires diffusion models that account for heterogeneity and crowding.

We have shown that protein diffusion in crowded membranes displays anomalous transport, manifested by subdiffusive behavior of the mean-square displacement.
The effect of crowding can be studied systematically with FCS
in a model lipid bilayer membrane where the concentration of avidin as crowding agent
is well controlled.
Depending on the surface coverage, we observe normal or anomalous transport separated by a transition regime.

We demonstrate an alternative way to analyze the FCS autocorrelation function by direct  conversion to the MSD. This approach is independent of the employed diffusion model and is even capable of identifing a possible temporal crossover from subdiffusive to diffusive motion.
The obtained MSDs, however, exhibit single power-law behavior at time scales between 1\,ms and 100\,ms for all concentrations, and we conclude that normal diffusion would take place at longer times beyond the scope of the present set of experiments.
A complementary approach would be to vary the beam waist of the laser, enabling a larger area of the bilayer to be measured and shifting the sensitivity of the experiment to longer times, which could elucidate a crossover mechanism from subdiffusive to diffusive motion.

The transition regime appears to be a mixed phase of micron-sized domains of normal diffusion coexisting with domains of subdiffusion.
Repeated measurements on the same spot  yielded reproducible results even after hours, showing that
the domains are stable and  long-living spatial  heterogeneities exist on top of the membrane; in particular we exclude aging effects.
The emergence of these spatial heterogeneities correlates with the onset of anomalous transport.
The latter is quantified in terms of the exponent of subdiffusion $\alpha$, which shows a continuous decrease as the degree of crowding is increased beyond the transition regime.
The mean-square displacements were extracted directly from the FCS autocorrelated data, independent of \emph{a priori} models for the motion. Then different transport mechanisms can be distinguished immediately by visual inspection of the MSD.

% We expect that avidin transport becomes normal at long time scales, but the crossover times in our model membrane were beyond the scope of our experiments.

A well-established mathematical model for subdiffusion is  the continuous time random walk (CTRW), where space is homogeneous and a broad distribution of waiting times is assumed from the very beginning.
In this model, one expects weak ergodicity breaking, i.e., time-averaged measurements exhibit aging~\cite{Lubelski:2008}. Our experiments are incompatible with this scenario; rather, they suggest that the anomalous transport originates from the observed spatial heterogeneities, similar to the phenomenology of transport on percolating systems~\cite{benAvraham:DiffusionInFractals}. An analogous conclusion
has been made recently for a three-dimensional crowded dextran solution~\cite{Szymanski:2009}.  From our experiments we exclude fractional Brownian motion (FBM)~\cite{Sebastian:1995} too as a valid description, since it predicts spatially homogeneous dynamics.

One may speculate on the nature of  the observed  heterogeneous structures.
The surface oligosaccharide groups of avidin as well as its
positive surface charge has been postulated to cause lateral gel-like network formation~\cite{Lou:2007}.
Therefore, it is possible that protein-protein interactions affect the spatial
organization of avidin at the transition and give rise to the observed heterogeneity.
The two-dimensional space accessible to membrane proteins may efficiently be blocked for topological reasons even at relatively low concentrations of proteins.
Thus, the onset of anomalous diffusion being at the low surface coverage of $5\%$ supports the idea that protein motion is obstructed by spacious, ramified structures.

In the context of cell membranes, another important question is how crowding affects the state of the \emph{lipids.} Here, we have taken advantage of fluid membrane that remains fluid even under a very crowded protein layer to eliminate this effect, and anomalous diffusion is due to protein crowding alone. The case of integral membrane proteins, such as rhodopsin, is more complex~\cite{Niu:2005}, and further studies are desirable, which introduce protein-induced heterogeneities in the lipid bilayer.

\section*{Acknowledgments}

Financial support from the German Excellence Initiative via the program
``Nanosystems Initiative Munich (NIM)''
and from the Deutsche Forschungsgemeinschaft (DFG) contract No. \mbox{FR 850/6-1}
is gratefully acknowledged.

% \bibliographystyle{rsc}
% \bibliography{fcs_avidin}

\end{document}